      [1999/12/01 v1.4c Il Nuovo Cimento]
\documentclass{cimento}
\usepackage{graphicx}  

\newcommand{\lesssim}{\lower.5ex\hbox{$\; \buildrel < \over\sim \;$}}
\newcommand{\gtrsim}{\lower.5ex\hbox{$\; \buildrel > \over\sim \;$}}
\newcommand{\swift}{{\it Swift}\ }
\newcommand{\apj}{{\it Astrophys.\ J.}}
\newcommand{\apjl}{{\it Astrophys.\ J.\ L.}}

\newcommand{\mnras}{{\it Monthly Not.\ Roy.\ Astron.\ Soc.}}

\title{The flat decay phase in the early X-ray afterglows of {\it Swift} GRBs}
\author{Jonathan Granot}
\instlist{
\inst
Kavli Institute for Particle Astrophysics and Cosmology, Stanford
University, P.O. Box 20450, MS 29, Stanford, CA 94309, USA;
granot@slac.stanford.edu}
\PACSes{\PACSit{98.70.Rz}{gamma--ray sources; gamma--ray bursts}}
\begin{document}

\maketitle

\begin{abstract}
Many \swift GRBs show an early phase of shallow decay in their X-ray
afterglows, lasting from $t\sim 10^{2.5}\;$s to $\sim 10^4\;$s after
the GRB, where the flux decays as $\sim t^{-0.2}-t^{-0.8}$.  This is
perhaps the most mysterious of the new features discovered by
\swift in the early X-ray afterglow, since it is still not clear what
causes it. I discuss different possible explanations for this
surprising new discovery, as well as their potential implications for
the gamma-ray efficiency, the afterglow kinetic energy, and perhaps
even for the physics of collisionless relativistic shocks.
\end{abstract}

\section{Introduction}
\label{sec:intro}

Before the launch of {\it Swift}, the monitoring of GRB afterglows
typically started at least several hours after the GRB. The
extrapolation back in time of the observed X-ray afterglow power law
flux decay usually gave a flux similar to that of the prompt emission
around the end of the GRB. Therefore, most people believed that \swift
would detect a simple single power law flux decay all the way from the
end of the prompt emission up to the late times that were observed
before \swift.  It had even been hoped that this would significantly
improve the constraints on the external density profile around the
GRB.

After the launch of \swift, however, a new and surprising picture soon
emerged, where the early X-ray afterglow showed several interesting
and unexpected features \cite{Nousek06}. These included mainly (i) an
initial rapid decay phase where $F_\nu \propto t^{-\alpha}$ with $3
\lesssim \alpha_1 \lesssim 5$ lasting from the end of the prompt
emission up to $\sim 10^{2.5}\;$s, (ii) a subsequent flat decay phase
where $0.2 \lesssim \alpha_2 \lesssim 0.8$, lasting up to $\sim
10^4\;$s (followed by the familiar pre-\swift power law decay with $1
\lesssim \alpha_3 \lesssim 1.5$), and (iii) X-ray flares, which appear
to be overlaid on top of the underlying power law decay in stages (i)
and (ii). The initial rapid decay stage appears to be a smooth
extension of the prompt emission \cite{Obrien06}, and is therefore
most likely the tail of the prompt GRB, probably due to emission from
large angles relative to our line of sight \cite{KP00}. The X-ray
flares appear to be a distinct emission component, as suggested by
their generally different spectrum compared to the underlying power
law component, and by the fact that the flux after a flare is usually
the continuation of the same underlying power law component from
before the flare \cite{Burrows05,Falcone06}. In many cases these
flares show sharp large amplitude flux variation on time scales
$\Delta t \ll t$ (see, e.g. \cite{Krimm06}), which are very hard to
produce by the external shock, and suggest a sporadic late time
activity of the central source.

The flat (or shallow) decay phase, stage (ii), and the initial rapid
decay phase, stage (i), appear to arise from two physically distinct
emission regions. This is supported by a change in the spectral index
that is observed in some of the transitions between these two stages
\cite{Nousek06}. Furthermore, the flat decay phase eventually smoothly
steepens into the familiar pre-\swift flux decay, which is well
established to be afterglow emission from the forward shock, strongly
suggesting\footnote{The reason for this is that obtaining a smooth
transition where the flux decay {\it steepens} between two distinct
emission regions requires fine tuning, and therefore is highly
unlikely to happen for practically every flat decay phase, as is
implied by observations.} that the flat decay phase is similarly
afterglow emission from the forward shock. This is also supported by
the fact that there is no evidence for a change in the spectral index
across this break \cite{Nousek06}.  Nevertheless, it is still not
clear what causes this shallow decay phase. Below I briefly describe
different possibilities and mention some of their possible
implications.

\section{Energy injection into the afterglow shock}
\label{sec:injection}

Perhaps the simplest explanation for the flat decay phase is gradual
and continuous energy injection into the afterglow (forward) shock
\cite{Nousek06,Panaitescu06,Zhang06}. This can take place in two main
forms \cite{Nousek06}: (1) a smooth distribution of ejected mass as a
function of its Lorentz factor, $M(>\Gamma) \propto \Gamma^{-s}$, and
its corresponding energy, $E(>\Gamma) \propto \Gamma^{-a}$ where $a =
s-1$. In this picture $\Gamma$ increases with radius
$R$.\footnote{This can naturally occur if toward the end of the prompt
GRB the Lorentz factor of the outflow that is being ejected decreases
with time. Even if $\Gamma$ does not monotonically increase with $R$
initially, but there is still some distribution in the initial Lorentz
factor, such an ordering (where $\Gamma$ increases with $R$) will
naturally be achieved as the result of internal shocks.} Material with
Lorentz factor $\Gamma$ catches up with the forward shock when the
Lorentz factor of the forward shock, $\Gamma_f$ , drops slightly below
$\Gamma$ \cite{RM98,SM00,R-R01}, resulting in a smooth and gradual
energy injection into the afterglow shock. (2) An alternative scenario
for the energy injection is that the central source remains active for
a long time
\cite{DL98,RM00,LR-R02,Dai04,R-R04}, where the ejected outflow (or
wind) has a Lorentz factor, $\Gamma_i \gg \Gamma_f$. This leads to a
highly relativistic reverse shock (with a Lorentz factor $\Gamma_r
\sim \Gamma_i/2\Gamma_f \gg 1$), while in scenario (1) the reverse
shock is only mildly relativistic, thus resulting in a different
emission from the reverse shock which may potentially be used in order
to distinguish between these scenarios.

\begin{figure}
\vspace{-0.4cm}
\hspace{-0.48cm}
\includegraphics[width=34.6pc]{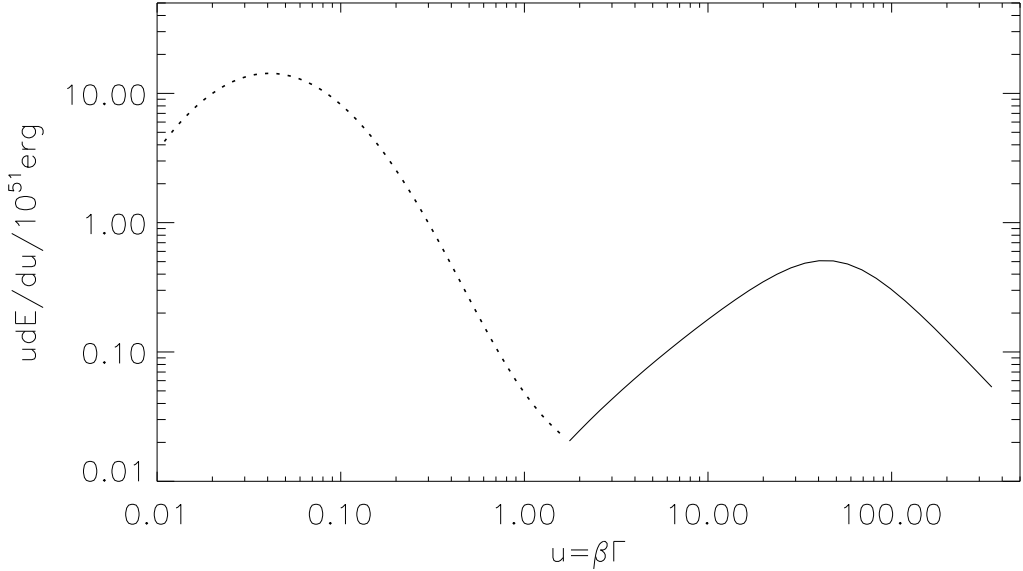}
\caption{\label{type1_inj}Schematic figure showing the distribution of
energy $E$ as a function of four velocity
$u\equiv\Gamma\beta=(\Gamma^2-1)^{1/2}$ ($dE/d\ln u$, in units of
$10^{51}$ erg), that is implied if the flat decay phase is due to type
(1) energy injection (from \cite{GK06}). It has one relativistic
component ({\it solid line}) with total energy $\sim 10^{51}\;$erg and
peak at $u\sim 30-50$ that produces the GRB ($u \gtrsim 10^2$) and the
afterglow radiations. The power-law index above the peak for this
component is well constrained by the X-ray data (the shallow part of
the light-curve) and is $dE/d\ln u \propto u^{-a}$ with $1 \lesssim a
\lesssim 2.5$ for a uniform external medium. The slope below the
peak is not well constrained (either than being positive) and is taken
to be 1. The second component (dashed curve) shows schematically the
kinetic energy in non-relativistic ejecta in the supernova
accompanying the GRB, which peaks at a typical velocity around $\sim
10^4\;{\rm km\; s^{-1}}$, and has an energy of the order of $\sim
10^{52}\;$erg.}
\end{figure}

In scenario (1) the observed shallow decay phases typically imply $1
\lesssim a \lesssim 2.5$ for a uniform external density and $a \gtrsim
5$ for a wind-like external density \cite{Nousek06,GK06} (which drops
as the inverse square of the distance from the source), while most of
the energy in the relativistic outflow is in material with $\Gamma
\sim 15 - 50$ for a uniform external density and $\Gamma \sim 10-20$
for a wind-like environment (see Fig.~\ref{type1_inj}). In scenario
(2) the observations typically imply an isotropic equivalent late time
energy deposition rate into the outflow of $L_{\rm iso} \propto t_{\rm
lab}^q$ with $q \sim -0.5$, where $t_{\rm lab}$ is the lab frame time
\cite{Nousek06}. The latter may, at least in some cases, be consistent
with the expectations for the spin-down luminosity of a newly born
millisecond magnetar \cite{Usov92,DL98} where $q$ varies smoothly from
$q \approx 0$ at early times (before considerable spin-down has
occurred) to $q \approx -2$ at late times.

\section{Viewing angles slightly outside the emitting region}
\label{sec:viewing}

\begin{figure}
\includegraphics[width=32.1pc]{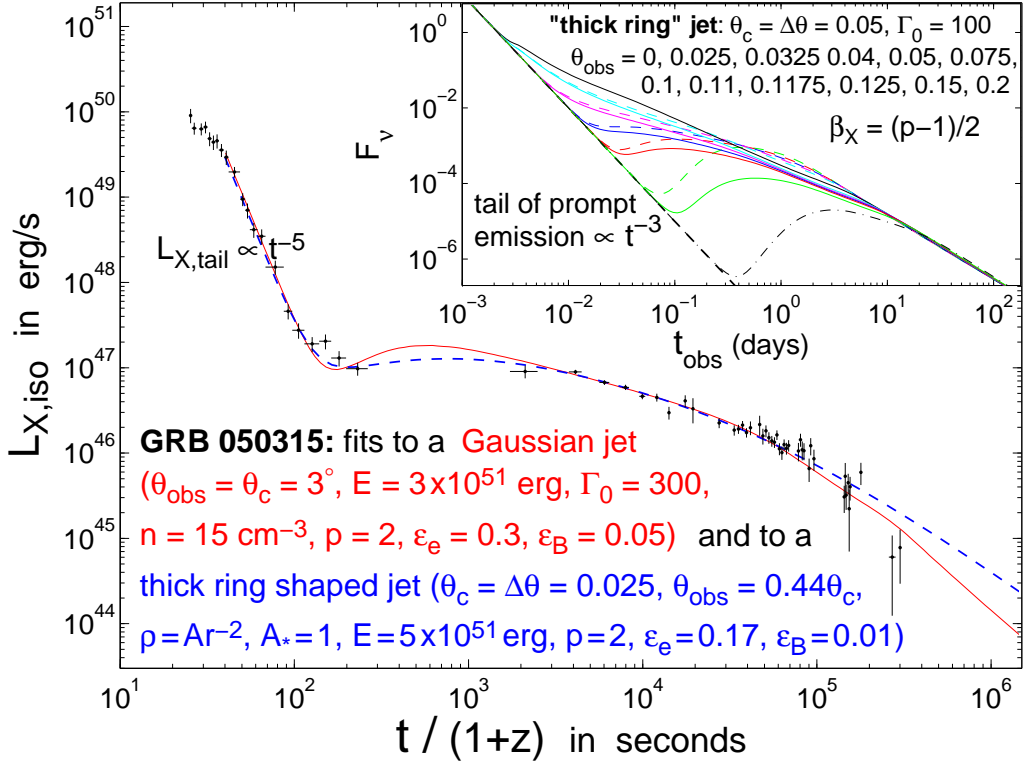}
\caption{\label{view}Tentative fits to the X-ray light curve 
of GRB~050315, for a viewing angle slightly outside the region of
bright afterglow emission (from \cite{EG06}). The inset demonstrates
the variety of different light curve shapes that are obtained for
different viewing angle, which may in principle accommodate the
observed diversity. For details see \cite{EG06}.}
\end{figure}

An interesting alternative explanation for the flat decay phase is a
viewing angle slightly outside the region of prominent afterglow
emission \cite{EG06}. In this interpretation the shallow decay phase
is the combination of the decaying tail of the prompt emission
(e.g. \cite{KP00}) and the gradual delayed onset of the afterglow for
such off-beam viewing angles (e.g. \cite{Granot02}), as is illustrated
in Fig.~\ref{view}. The fact that such a flat decay phase is observed
in a large fraction of \swift X-ray afterglows, in most of which the
prompt $\gamma$-ray emission is relatively bright, suggests in this
picture that many observers have a very high ratio of $\gamma$-ray to
kinetic (isotropic equivalent) energy at early times along their line
of sight. This requires a high efficiency of the prompt $\gamma$-ray
emission, of $\epsilon_\gamma
\gtrsim 90\%$, under the standard assumptions of afterglow theory, as
discussed in \S~\ref{sec:gamma_eff}.

\section{Two component jet}
\label{sec:2comp}

This explanation envisions a distribution of the initial Lorentz
factor as a function of direction, i.e. with angle within the
collimated outflow, $\Gamma_0 = \Gamma_0(\theta)$. This should not be
confused with the distribution in the initial Lorentz factor along the
same direction, as in type (1) energy injection that was discussed in
\S~\ref{sec:injection}. The local deceleration time (at which most of
the local energy is transferred to the shocked external medium) at
each point in the jet depends sensitively on the local value of the
initial Lorentz factor (and less sensitively on the local energy per
solid angle; see, e.g., Eq.~12 of
\cite{GK03}). Therefore, the regions that contribute to the prompt
$\gamma$-ray emission, which typically have an initial Lorentz factor
$\Gamma_0 \gtrsim 10^2$, decelerate early on (on a time scale similar
to the duration of the GRB), while region with smaller $\Gamma_0$
decelerate and start contributing significantly to the afterglow
emission only at later times. 

\begin{figure}
\includegraphics[width=32.1pc]{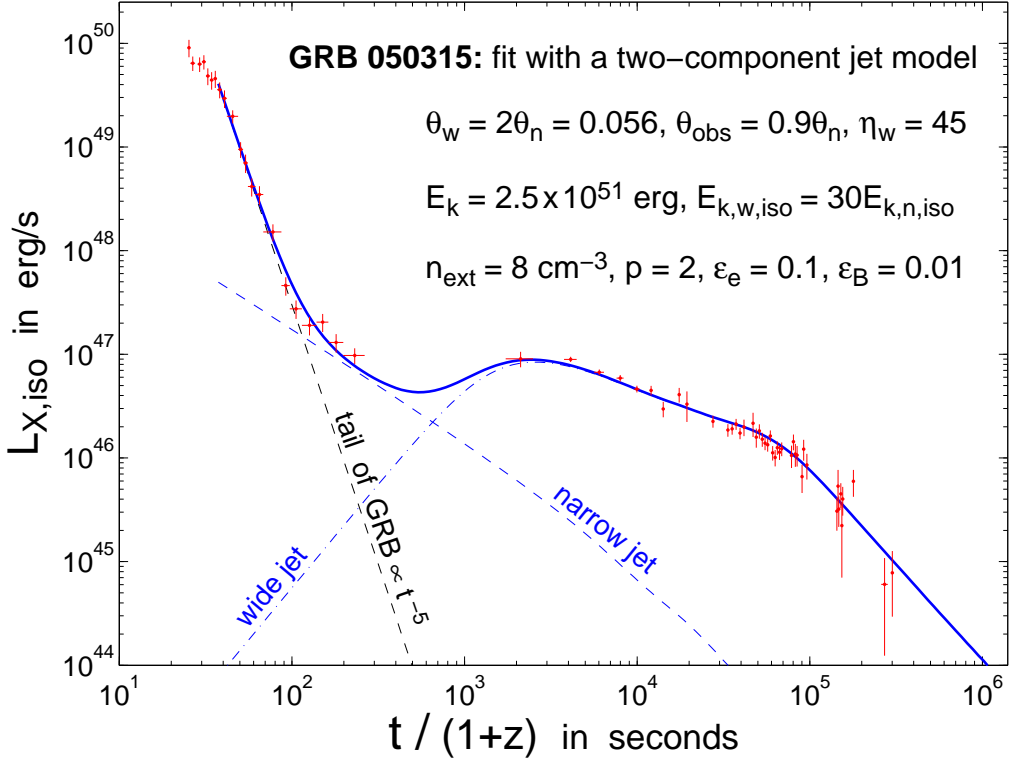}
\caption{\label{2comp}Tentative fit to the X-ray light curve 
of GRB~050315, for a two component jet (from
\cite{GKP06}). For details see the text and/or \cite{GKP06}.}
\end{figure}

In the simplest version of this picture there are two discrete jet
components: an inner narrow jet of half-opening angle $\theta_n$ with
$\Gamma_0 = \eta_n \gtrsim 10^2$, surrounded by a wider jet of
half-opening angle $\theta_w > \theta_n$ with a smaller $\Gamma_0 = \eta_w \sim
10-30$. Theoretical motivation for such a jet structure has been found
both in the context of the cocoon in the collapsar model
\cite{R-RCR02} and in the context of a hydromagnetically driven
neutron-rich jet \cite{VPK03}. It was invoked by \cite{PKG05} as a
possible way to alleviate the pre-\swift constraints on the
$\gamma$-ray emission efficiency, $\epsilon_\gamma$. However, \swift
observations show that while it may produce the observed flat decay
phase (as is illustrated in Fig.~\ref{2comp}), it still requires a
very high $\epsilon_\gamma$, under standard assumptions \cite{GKP06}.

\section{Initial increase with time of the afterglow efficiency}
\label{sec:AG_eff}

All previous explanations relied on an increase with time in the
typical afterglow isotropic equivalent kinetic energy, $E_{\rm
k,iso}$, within the observed region, either through an increase in
$E_{\rm k,iso}$ along the line of sight (\S~\ref{sec:injection}) or as
other regions start to contribute significantly to the observed flux
(\S\S~\ref{sec:viewing},\ref{sec:2comp}). In this explanation,
however, $E_{\rm k,iso}$ in the observed region remains constant from
very early on (around the deceleration time), but the efficiency of
the X-ray afterglow emission, $\epsilon_X$, initially increases with
time \cite{GKP06}.

It is natural to define $\epsilon_X(t) \equiv tL_{X,{\rm iso}}/E_{\rm
k,iso}$ where $L_{X,{\rm iso}}$ is the afterglow isotropic equivalent
X-ray luminosity. The ratio $\epsilon_X(t)E_{\rm k,iso}(t)/tF_X(t)$
(where $F_X$ is the X-ray afterglow flux) depends only on the
redshift, and is constant in time (see Eq.~15 of \cite{GKP06}). In the
flat decay phase $F_X \propto t^{-\alpha}$ with $\alpha_2 < 1$, so
that $tF_X(t)$ increases with time, and therefore $\epsilon_X(t)E_{\rm
k,iso}(t)$ must similarly increase with time. In other explanations
this was attributed to an increase with time of $E_{\rm k,iso}(t)$
(where $\epsilon_X(t)$ slowly decreases with time), while in the
current explanation $E_{\rm k,iso}$ remains constant, while
$\epsilon_X$ initially increases with time, thus causing the shallow
decay phase. This might be causes by a value of $p < 2$ for the power
law index of the electron energy distribution
\cite{GKP06}. In this case, however, radiative losses might become
important (which would steepen the flux decay) and this option is
often inconsistent with the measured spectral slope in the X-rays.

A more interesting way for $\epsilon_X$ to increase with time is due
to an increase with time in one or more of the following shock
microphysical parameters: the fraction of the internal energy in
relativistic electrons, $\epsilon_e$, or in magnetic fields,
$\epsilon_B$, or the fraction $\xi_e$ of the electrons that are
accelerated to a relativistic power-law distribution of energies.
(for more details see \cite{GKP06}). In this scenario, the shock
microphysical parameters eventually saturate at some asymptotic
values, bringing the flat decay phase to an end. If this is indeed the
cause for the shallow decay phase, then the observations of this phase
can potentially be used in order to constrain the physics of
collisionless relativistic shocks.

\section{Implications for the gamma-ray efficiency and for the jet energy}
\label{sec:gamma_eff}

Pre-{\it Swift} studies \cite{PK02,Yost03,L-RZ04} found that $E_{\rm
k,iso}$ at late times (typically evaluated at $t_* = 10\;$hr), $E_{\rm
k,iso}(t_*)$, is comparable to the isotropic equivalent energy output
in $\gamma$-rays, $E_{\rm \gamma,iso}$, i.e. that typically $\kappa
\equiv E_{\rm \gamma,iso}/E_{\rm k,iso}(t_*) \sim 1$. The $\gamma$-ray
efficiency is given by $\epsilon_\gamma = E_{\rm
\gamma,iso}/(E_{\rm \gamma,iso}+E_{\rm k,iso,0})$, where $E_{\rm
k,iso,0}$ is the initial value of $E_{\rm k,iso}$ corresponding to
material with a sufficiently large initial Lorentz factor ($\Gamma_0
\gtrsim 10^2$) that could have contributed to the $\gamma$-ray
emission. This implies a simple relation,
$\epsilon_\gamma/(1-\epsilon_\gamma) = \kappa f$, where $f \equiv
E_{\rm k,iso}(t_*)/E_{\rm k,iso,0}$ can be estimated from the
early afterglow light curve \cite{GKP06}.

Interpreting the shallow decay phase in the early X-ray afterglow as
energy injection \cite{Nousek06,Zhang06,Panaitescu06,GK06} typically
implies $f \gtrsim 10$ and $\epsilon_\gamma \gtrsim 0.9$
\cite{Nousek06,EG06,GKP06}. This is a very high efficiency for any
reasonable model for the prompt emission, and in particular for the
popular internal shocks model. If the shallow decay phase is not
caused by energy injection, but is instead due to an increase with
time in the afterglow efficiency, then $f \sim 1$ and typically
$\epsilon_\gamma \sim 0.5$ \cite{GKP06}.  This is a more reasonable
efficiency, but still rather high for internal shocks. If, in
addition, $E_{\rm k,iso}(10\;{\rm hr})$ had been underestimated,
e.g. due to the assumption that $\xi_e = 1$, then\footnote{\cite{EW05}
have pointed out a degeneracy where the same afterglow observations
are obtained under the substitution $(E,n) \to (E,n)/\xi_e$ and
$(\epsilon_e,\epsilon_B) \to
\xi_e(\epsilon_e,\epsilon_B)$ for a value of $\xi_e$ in the range
$m_e/m_p \leq \xi_e \leq 1$, instead of the usual assumption of $\xi_e
= 1$.} $\kappa \sim \xi_e$ and $\xi_e \sim 0.1$ would lead to $\kappa
\sim 0.1$ and $\epsilon_\gamma \sim 0.1$.

The internal shocks model can reasonably accommodate $\gamma$-ray
efficiencies of $\epsilon_\gamma \lesssim 0.1$, which in turn imply
$\kappa \lesssim 0.1$. Since the true (beaming-corrected) $\gamma$-ray
energy output, $E_\gamma = f_b E_{\rm\gamma,iso}$ (where $f_b \approx
\theta_0^2/2$ and $\theta_0$ is the half-opening angle of the uniform
jet), is clustered around $10^{51}\;$erg \cite{Frail01,BFK03}, this
implies $E_k(t_*) = f_b E_{\rm k,iso}(t_*) = \kappa^{-1}E_\gamma
\gtrsim 10^{52}\;$erg for a uniform jet.  For a structured jet with
equal energy per decade in the angle $\theta$ from the jet symmetry
axis ($dE/d\Omega \propto
\theta^{-2}$) in the wings (between some inner core angle $\theta_c$
and outer edge $\theta_{\rm max}$), the true energy in the jet is
larger by a factor of $1+2\ln(\theta_{\rm max}/\theta_c) \sim 10$,
which implies $E_k(t_*) \gtrsim 10^{53}\;$erg in order to achieve
$\epsilon_\gamma \lesssim 0.1$. Such energies are comparable (for a
uniform jet) or even higher (for the latter structured jet) than the
estimated kinetic energy of the Type Ic supernova (or hypernova) that
accompanies long-soft GRBs. This is very interesting for the total
energy budget of the explosion.

\section{Conclusions}

The flat (or shallow) decay phase is arguably the most striking of the
new features found by \swift in the early X-ray emission of
GRBs. Nevertheless, it is still not clear what causes it. There are
many possible explanations, which include energy injection into the
afterglow shock (\S~\ref{sec:injection}), viewing angle effects
(\S~\ref{sec:viewing}), a two component jet (\S~\ref{sec:2comp}), or
an initial increase with time in the efficiency of the afterglow
emission (\S~\ref{sec:AG_eff}). Good monitoring of the early afterglow
emission over a wide range in frequency, might help to distinguish
between the different explanations. It is not obvious, however,
whether any of the explanations that have been mentioned here is
indeed the dominant cause for the shallow decay phase. Evidence is
accumulating that the steepening in the flux decay at the end of the
shallow decay phase is chromatic \cite{Panaitescu06b} -- it is
observed in the X-rays but not in the optical. This is in
contradiction with the expectations of all explanations that have been
put forth so far for the flat decay phase, where the steepening at its
end is expected to be largely achromatic.  There is definitely still a
lot of work ahead of us in trying to understand the origin of these
fascinating new observations in the
\swift era.

\acknowledgments
I am grateful to Arieh K\"onigl, Tsvi Piran, Pawan Kumar, David
Eichler, Enrico Ramirez-Ruiz, and Chryssa Kouveliotou for their
collaboration on part of the work that is presented here. This
research was supported by the US Department of Energy under contract
number DEAC03- 76SF00515.

\end{document}